\documentclass[10pt,twocolumn]{article}

\usepackage[utf8]{inputenc}
\usepackage[T1]{fontenc}
\usepackage{amsmath,amssymb}
\usepackage{graphicx}
\usepackage{booktabs}
\usepackage{hyperref}
\usepackage{listings}
\usepackage{xcolor}
\usepackage[margin=0.85in]{geometry}
\usepackage{enumitem}
\usepackage{caption}
\usepackage{algorithm}
\usepackage{algpseudocode}
\usepackage{multirow}
\usepackage{tabularx}
\usepackage{fancyvrb}
\usepackage{tikz}
\usetikzlibrary{positioning,arrows.meta,fit,backgrounds}

\title{\textbf{Governed MCP: Kernel-Level Tool Governance\\for AI Agents via Logit-Based Safety Primitives}}

\author{
  Daeyeon Son\\
  Independent Researcher\\
  Republic of Korea\\
  \texttt{sdy1350@gmail.com}
}

\date{July 2026}

\begin{document}
\maketitle

\begin{abstract}
AI agents increasingly call external tools (file system, network, APIs)
through the Model Context Protocol (MCP)~\cite{mcp-spec}. These tool calls
are the agent's syscalls---privileged operations with side effects on
shared state---yet today's safety enforcement lives entirely in
userspace, where a 10-line script can bypass it. I propose
\textbf{Governed MCP}, a kernel-resident tool governance gateway built
on a logit-based safety primitive (ProbeLogits, companion paper). The
gateway interposes on every MCP tool call in a 6-layer pipeline:
schema validation, trust tier check, rate limit, adversarial pre-filter,
\emph{ProbeLogits gate} (the load-bearing semantic check), and
constitutional policy match, with a Blake3-hashed audit chain.

I implement Governed MCP in \textbf{Anima OS}, a bare-metal
\texttt{x86\_64} OS in $\sim$286{,}000 lines of Rust (378
files). The five non-inference layers and the audit append
add a measured \textbf{11.3\,$\mu$s} of overhead per call
(in-OS microbench, $n$=64 benign calls; v1's unarchived
65.3\,$\mu$s estimate is superseded by this ${\sim}$6$\times$
lower measured value, whose components sum consistently);
the ProbeLogits gate---one probe-prompt prefill plus a single
logit read---costs 358/332/556\,ms per classification on
Qwen2.5-7B~/ Llama~3~8B~/ Mistral~7B (hosted remeasurement),
2.4--3.4$\times$ faster than a Llama Guard 3 pass on the same
hardware. A silicon-measured ablation on a 101-prompt
MCP-domain benchmark shows that removing the
ProbeLogits layer collapses F1 from 0.789 to 0.357
($\Delta$F1 = $-$0.432)---hand-rule
firewalling alone is insufficient. A rebuilt in-OS mediation
harness enumerates 59 agent-to-system paths: all 18
WASM-to-system host functions (15 governed, 3 read-only) and
all 20 registered MCP tools are mediated by the kernel
gateway, while of the 21 ring-3 syscalls, 8 reach the
gateway, 7 touch no governed resource, and 6
resource-reaching paths remain ungated pending the M4.2
milestone---a disclosed limitation
(the scope and caveats of the complete-mediation claim are
stated in \S\ref{sec:design-mediation}); a 10-LoC userspace
bypass that defeats existing guardrail libraries is
structurally impossible against the kernel-resident gate.
New in this revision, multi-model validation across three
architectures (HarmBench 98--99\% non-copyright block rate,
XSTest 98.5--100\% unsafe recall, ToxicChat parity with
Llama Guard 3) shows that the substrate primitive the gateway
is built on is architecture-agnostic; these are
substrate-level benchmarks of the ProbeLogits primitive,
while the gateway's MCP-domain behavior is evaluated by the
ablation and mediation harness above.

Governed MCP demonstrates that tool-call
governance is feasible as an OS primitive, not just an
application-layer concern.
\end{abstract}

\section{Introduction}
\label{sec:intro}


AI agents are increasingly autonomous, executing multi-step
plans via tool calls (file reads, web fetches, code execution,
shell commands). The Model Context Protocol (MCP), introduced
by Anthropic~\cite{mcp-spec}, has become the de facto standard
for tool calling: a JSON-RPC interface where the agent
proposes a tool name and arguments, and a server executes them.

\textbf{Tool calls are syscalls.} Like syscalls, they cross the
privilege boundary between agent (less trusted) and host (more
trusted), and they carry the same risks: argument exfiltration,
side effects on shared state, escalation. Unlike syscalls,
they have no kernel mediation. Existing safety
infrastructure---NeMo Guardrails~\cite{nemoguardrails}, AGT
schemas~\cite{msagt}, AutoGPT-style~\cite{autogpt} wrappers---all
operate as Python libraries imported by the agent process,
in the same address space and with the same privilege as the
agent itself. \emph{The agent decides whether to call them.}

The structural problem is not that these libraries are weak,
but that they are libraries: they cannot be the privilege boundary
between the agent and the host. Listing~\ref{lst:bypass} shows
three increasingly subtle ways an agent (or compromised tool
that exfiltrated credentials) can bypass them, none of which
require a vulnerability in the safety library:

\begin{lstlisting}[float=t,caption={Three bypass paths for in-process safety libraries. None exploit a library bug; all rely on the agent process having the same privilege as the safety library.},label={lst:bypass},language=Python]
# (1) Skip the import. The library only runs if the
#     agent code chooses to call it.
from openai import OpenAI; from os import system
# no `from nemo_guardrails import LLMRails`
client = OpenAI()
plan = client.chat.completions.create(...)
system(plan.choices[0].message.content)  # unguarded

# (2) Replace the function. Same address space => same write
#     permission to the library's symbols.
import nemoguardrails
nemoguardrails.LLMRails.generate = lambda self, p: p
# now every "guarded" call returns the prompt unfiltered

# (3) Inject the safety verdict. Many libraries return a
#     dict; mutate it before the caller checks it.
result = guardrails.check(prompt)
result["allowed"] = True
# downstream gate sees allowed=True
\end{lstlisting}

These bypasses succeed because Python imports, attribute writes,
and dict mutations all run with the agent's own privilege. No
amount of library-level hardening (input validation, signed
checks, etc.) closes the gap, because the gap is at the
\emph{privilege boundary}: there is no boundary. Any defense
the library implements, the agent's process can also undo.

The same critique applies to JavaScript wrappers, Go middleware,
and any other in-process safety layer. Kernel-grade governance
requires that the safety check live in a privilege domain the
agent cannot enter.

\textbf{This paper proposes Governed MCP}, a kernel-resident
governance gateway that interposes on every MCP tool call. The
gateway runs in the OS kernel of Anima OS---a bare-metal
\texttt{x86\_64} operating system written in Rust. Agents execute as
WASM~\cite{wasmcore} bytecode in a sandbox above the gateway;
the only path from the sandbox to a tool call is through host
functions that the kernel exposes, every side-effect-producing
one of which routes through the gateway
(\S\ref{sec:design-mediation}). Because the agent's WASM context cannot
address the gateway's memory and cannot invoke kernel code
except through the WASM ABI, the agent \emph{cannot skip the
check by design}: there is no userspace shim to monkey-patch
because the check is not in userspace.

The gateway is structured as a six-layer pipeline. The first
four layers are syntactic and policy checks: JSON-RPC schema
validation, trust-tier whitelist (which tools may be invoked
by which trust class of agent), token-bucket rate limit, and
an O($n$) regex pre-filter for prompt-injection patterns. The
fifth layer is the load-bearing semantic check: a single
forward pass through the loaded inference model that reads
``Safe'' vs.\ ``Dangerous'' logits at the verbalizer position
(the ProbeLogits primitive). The sixth applies a 12-principle
constitutional policy match. Each tool call appends a
Blake3-hashed audit record. End-to-end cost is dominated by
the inference layer: one probe classification (prompt prefill
plus a single logit read) costs 358\,ms on Qwen2.5-7B Q4\_0
(332\,ms on Llama~3~8B, 556\,ms on Mistral~7B, hosted
remeasurement, \S\ref{sec:eval-perf}); the five non-inference
layers plus the audit append add only a measured 11.3\,$\mu$s of overhead
(\S\ref{sec:eval-perf}).

\paragraph{Why this matters now.}
The Model Context Protocol entered widespread deployment in
2024--2025, and major frameworks (Anthropic's Claude clients,
OpenAI's Tool API, Microsoft's Copilot tools) now ship MCP
support by default. This means tool-call governance is no
longer a research concern but a deployment one---and the
governance layer that ships in current production stacks is
exactly the in-process Python-library design that
Listing~\ref{lst:bypass} defeats. The window to define what a
proper kernel-grade MCP gateway should look like is open
\emph{now}, before deployment patterns harden around the
current weak architecture.

\paragraph{Contributions.}
\begin{enumerate}[leftmargin=*]
  \item \textbf{A 6-layer kernel-resident MCP governance
    pipeline} (schema, trust, rate, adversarial pre-filter,
    ProbeLogits, constitutional) with a measured
    11.3\,$\mu$s non-inference overhead (\S\ref{sec:eval-perf}) and
    FAIL-CLOSED semantics: if the
    inference engine is unavailable, all tool calls are denied.
  \item \textbf{Demonstration that the ProbeLogits semantic
    layer is load-bearing}: a 4-configuration ablation on a
    101-prompt MCP-domain benchmark (37 dangerous + 64 benign;
    the corpus is in-tree and verified), now silicon-measured
    in-OS, shows that removing the ProbeLogits gate collapses
    F1 from 0.789 to 0.357 ($\Delta$F1 = $-$0.432)
    (\S\ref{sec:eval-ablation}). Hand-rule firewalling without
    semantic understanding is insufficient.
  \item \textbf{Complete mediation of the WASM ABI surface,
    enforced structurally}: every side-effect-producing
    WASM-to-system host function (18 registered: 15 base + 3
    MCP) is mediated by the kernel gateway---the base
    functions via the \texttt{governance\_check\_host()}
    choke-point and the MCP tool-call entry via the gateway
    pipeline itself---and the gateway's inner MCP entry point
    is \texttt{pub(crate)} (not in the WASM ABI). An in-OS
    test suite (\texttt{webtest mcp}) exercises the MCP
    surface end-to-end against the gateway's audit counters,
    and a rebuilt enumerated-path harness (silicon-measured,
    single boot) covers 59 paths: 18 host functions (15
    governed + 3 read-only), 20 MCP tools (all
    verdict-verified via dry-run evaluation), and 21 ring-3
    syscalls (7 gateway-governed + 1 self-gated + 2 read-only
    metadata + 5 pure), with \textbf{6 resource-reaching
    ring-3 syscalls ungated and deferred to milestone
    M4.2}---a disclosed limitation that replaces the v1
    ``123 paths, 100\% mediated'' claim.
    The argument is scoped to the WASM ABI surface (not the
    entire kernel; out-of-scope items are listed in
    \S\ref{sec:threat}).
  \item \textbf{Open-source bare-metal implementation}
    ($\sim$286{,}000 lines of Rust across 378 files,
    AGPL-3.0; the bare-metal kernel alone is ${\sim}$194K
    lines) demonstrating that
    kernel-grade governance is feasible without proprietary
    silicon or hypervisor tricks.
\end{enumerate}

The remainder of the paper is organized as follows.
\S\ref{sec:bg} surveys the MCP protocol and existing safety
infrastructure. \S\ref{sec:threat} fixes the threat model.
\S\ref{sec:design} describes the gateway's six-layer design
and the ProbeLogits substrate. \S\ref{sec:eval} evaluates the
gateway end-to-end. \S\ref{sec:disc} discusses generalization,
limitations, and future work; \S\ref{sec:related} positions the
work; \S\ref{sec:conclusion} concludes.

\section{Background}
\label{sec:bg}

\paragraph{The Model Context Protocol.}
MCP~\cite{mcp-spec}, introduced by Anthropic in late 2024, is
a JSON-RPC~\cite{jsonrpc} interface that standardizes how an
LLM-driven agent discovers and invokes external tools. A
\emph{tool} is a function described by a JSON Schema for its
input arguments and output shape; tools are hosted by an MCP
\emph{server} (a Python or Node process bound to a transport)
and consumed by an MCP \emph{host} (the agent runtime).
The protocol defines four primary methods:
\texttt{list\_tools} (server returns its catalog of tool
descriptions), \texttt{call\_tool} (host invokes a tool with
JSON arguments and receives a JSON or stream result),
\texttt{list\_prompts}, and \texttt{list\_resources}. Two
transports are supported: stdio (in-process subprocess) and
Streamable HTTP (long-polling over HTTP/1.1 or HTTP/2).

\paragraph{Why MCP is the right abstraction.}
Prior to MCP, every agent framework defined its own tool
interface: LangChain \texttt{Tool}~\cite{langchain}, OpenAI
\texttt{function\_call} schema~\cite{openai-fc}, and
proprietary AGT~\cite{msagt} all required tool authors to
write framework-specific adapters. MCP factors out the tool
catalog and invocation protocol, so a single MCP server can
be consumed by Claude clients, OpenAI clients, AutoGPT-style
agents, and custom runtimes without modification. This
factoring also makes MCP the right \emph{governance}
chokepoint: every tool invocation, regardless of which agent
framework called it, traverses the same JSON-RPC method, so a
single gateway can mediate all of them. Enforcement at the
protocol layer dominates enforcement at the framework layer
because it cannot be bypassed by switching frameworks.

\paragraph{Where MCP servers run today.}
In current production deployments, MCP servers run as
unprivileged userspace processes on the developer's machine
(stdio transport) or as remote services accessed over HTTP.
The MCP host (the agent runtime) is also a userspace process,
typically a Python application using the
\texttt{mcp-server-stdio} or \texttt{mcp-client-http}
libraries. Safety enforcement is left to the host's own code;
nothing in the MCP specification requires the host to call a
safety check before invoking a tool, and nothing in the OS
prevents an agent that controls the host process from skipping
the check. The check and the agent share a UID, an address
space, and a Python interpreter. This is the deployment
pattern that the kernel-resident gateway proposed in this
paper replaces.

\paragraph{Existing safety enforcement.}
Three families of system are deployed today:
\textbf{(1) Fine-tuned classifiers}---Llama Guard~3~\cite{llamaguard3},
WildGuard~\cite{wildguard}---add a second model invocation per
request. Latency is high (${\sim}$1\,s per check), and the
classifier model itself runs in the same userspace as the agent.
\textbf{(2) Schema-only validators}---AGT~\cite{msagt}, JSON
Schema---block syntactic violations but cannot reason about
semantics. ``Read this file'' passes schema; ``read /etc/shadow''
also passes.
\textbf{(3) Wrapper libraries}---NeMo Guardrails~\cite{nemoguardrails}
runs Python rule chains around model calls. The library is
imported by the agent process; one line of code disables it
(Listing~\ref{lst:bypass}).

None of these enforce \emph{below} the agent's privilege boundary.

\section{Threat Model}
\label{sec:threat}

\paragraph{Attacker.}
The attacker is the AI agent itself. The threat is realistic
without assuming the agent is malicious by design: an agent
trained on, fine-tuned with, or prompted by adversarial input
can produce dangerous tool-call sequences while running
ostensibly benign code. Concretely, the attacker manifests in
five ways the gateway must defend against:

\begin{enumerate}[leftmargin=*,topsep=2pt,itemsep=1pt]
  \item \textbf{Direct adversarial agent.} The agent's prompt
    or fine-tuning has been compromised (e.g., a malicious
    system prompt or a poisoned training dataset). It actively
    attempts dangerous tool calls.
  \item \textbf{Indirect prompt injection.} A tool's output
    (web page, file content, API response) contains
    adversarial instructions that the agent's next reasoning
    step incorporates as if they were system instructions.
    This is the agent attacker class identified by Greshake
    \emph{et al.}~\cite{greshake2023} and confirmed in the
    field by Apple, Slack, and Notion incidents in 2024--2025.
  \item \textbf{Jailbreak / role-play.} The agent is asked
    (e.g., by an end user) to ``play the role of'' an
    unconstrained assistant and execute the resulting plan.
    Refusal training is bypassed by the role frame; tool
    calls follow.
  \item \textbf{Multi-step decomposition.} Each individual
    tool call appears benign in isolation, but their
    composition causes harm
    (e.g., \texttt{read /etc/passwd} $\to$ \texttt{web\_post}
    $\to$ exfiltration). Per-call governance must reason
    about call sequences as well as individual calls.
  \item \textbf{Compromised MCP server.} A third-party MCP
    server returns malicious results that, when fed back
    into the agent's context, produce indirect injection
    (case 2). The gateway must treat tool outputs as
    untrusted input.
\end{enumerate}

\paragraph{Trust boundary.}
The gateway runs in kernel ring 0; the WASM sandbox runs at
the WASM/Cranelift virtualization layer with no \texttt{x86\_64}
syscall capability and no ability to address gateway memory
(this claim assumes a correct Cranelift implementation;
JIT-bug bypass classes are discussed in
\S\ref{sec:design-mediation}, ``WASM isolation soundness'').
The agent can invoke the gateway only via the
\texttt{governance\_check\_host()} host function, which is the
single mediated entry. This is the standard reference monitor
position~\cite{anderson1972,saltzer1975}: the agent's only
egress to system effects routes through one place, and that
place runs at higher privilege than the agent.

\paragraph{In scope.}
(i) tool-call arguments that exfiltrate sensitive data,
(ii) indirect prompt injection that escalates to dangerous
tool calls, (iii) jailbreak/role-play prompts that attempt
to override system instructions, (iv) chained multi-step tool
sequences that hide intent in benign-looking individual calls,
(v) compromised MCP server outputs as injection vector,
(vi) FAIL-CLOSED guarantee under inference engine failure.

\paragraph{Out of scope.}
Side channels via tool latency or argument size (the gateway
does not pad timing); kernel exploits in Anima OS itself
(which would compromise the gateway directly); hardware-level
tampering (Rowhammer, voltage glitching, etc.);
adversarial inference models (a maliciously trained model
loaded into the kernel would invalidate the substrate);
\emph{time-of-check-to-time-of-use (TOCTOU) races on tool
arguments} (e.g., a path resolved by the tool after the
gateway evaluated the argument string---in single-threaded
agent execution this is closed, in concurrent execution it is
not); and \emph{cross-agent collusion via shared memory
regions} (the gateway checks per-call intent on the control
plane but does not inspect the data plane).
Defense against these requires complementary mechanisms
(constant-time padding, formal kernel verification,
attested boot, model provenance, per-tool argument capture,
data-flow tracking) that are orthogonal to the gateway design.

\section{Design \& Implementation}
\label{sec:design}


\subsection{System Architecture Overview}
\label{sec:design-arch}

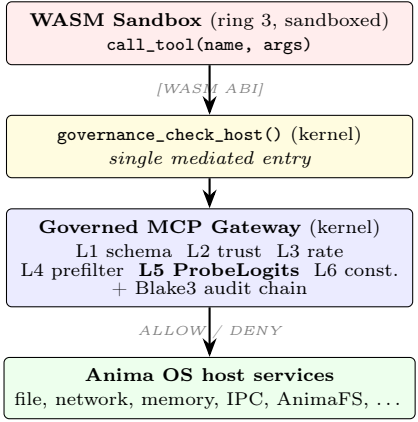
\begin{figure}[t]
\centering
\begin{tikzpicture}[
  font=\scriptsize,
  every node/.style={rectangle, rounded corners=2pt, draw, align=center, minimum width=5.4cm, inner sep=4pt},
  arrow/.style={-{Stealth[length=2.5mm]}, thick},
  level/.style={font=\tiny\itshape, gray}
]
\node[fill=red!7] (sandbox) {\textbf{WASM Sandbox} (ring 3, sandboxed)\\[1pt]
{\scriptsize\ttfamily call\_tool(name, args)}};

\node[level, draw=none, below=2mm of sandbox, fill=none, minimum width=0pt, inner sep=1pt] (abi) {[WASM ABI]};

\node[fill=yellow!15, below=2mm of abi] (gck)
  {\texttt{governance\_check\_host()} (kernel)\\[1pt]
   {\scriptsize\itshape single mediated entry}};

\node[fill=blue!8, below=4mm of gck, minimum width=5.4cm] (gw)
  {\textbf{Governed MCP Gateway} (kernel)\\[1pt]
   {\scriptsize L1 schema\ \ L2 trust\ \ L3 rate}\\[-1pt]
   {\scriptsize L4 prefilter\ \ \textbf{L5 ProbeLogits}\ \ L6 const.}\\[-1pt]
   {\scriptsize + Blake3 audit chain}};

\node[level, draw=none, below=2mm of gw, fill=none, minimum width=0pt, inner sep=1pt] (verd) {ALLOW / DENY};

\node[fill=green!7, below=2mm of verd] (host)
  {\textbf{Anima OS host services}\\[1pt]
   {\scriptsize file, network, memory, IPC, AnimaFS, \ldots}};

\draw[arrow] (sandbox) -- (gck);
\draw[arrow] (gck) -- (gw);
\draw[arrow] (gw) -- (host);
\end{tikzpicture}
\caption{Trust-boundary placement of the gateway.
The WASM agent's only egress to system effects is the
\texttt{governance\_check\_host()} entry, which routes through
all six layers before any host service is invoked.}
\label{fig:architecture}
\end{figure}

Figure~\ref{fig:architecture} shows the gateway's position
between the WASM agent and the OS host services. The agent
cannot address gateway memory and cannot invoke host services
except through the mediated entry; the gateway runs at full
kernel privilege.

\subsection{Six-Layer Pipeline}
\label{sec:design-pipeline}

Every MCP \texttt{call\_tool} request traverses six layers in
fixed order (Table~\ref{tab:layers}). The first four are
syntactic and policy checks, fast enough to reject obviously
invalid requests without any LLM cost. The fifth---the
ProbeLogits gate---is the only semantic check and the only
layer that requires an LLM forward pass. The sixth applies
12-principle constitutional policy matching. An audit record
is written via Blake3 hash chain for every decision.

\paragraph{Layer 1 (schema validation).}
JSON-RPC parsing followed by MCP tool-spec match: the
incoming \texttt{call\_tool} request is parsed into a
\texttt{Request} struct and matched against the tool's
declared input JSON Schema. Type mismatches, missing required
fields, and unknown tool names deny here. This layer never
performs network or filesystem I/O; cost is bounded by the
size of the JSON payload (typically $<$1\,KB).

\paragraph{Layer 2 (trust tier).}
Anima OS classifies agents into four trust tiers
(\texttt{System}, \texttt{AiNative}, \texttt{AiEnhanced},
\texttt{Classic}) based on origin, signature, and prior trust
evolution. Each tool declares its minimum required tier; the
gateway checks that the calling agent's tier is at least the
required level. A \texttt{Classic} agent (e.g., a
community-uploaded WASM with no verified provenance) cannot
invoke \texttt{system\_shell\_exec} regardless of arguments;
the layer denies before semantic analysis runs. This is a
purely policy-driven check, a measured 1.15\,$\mu$s per call.

\paragraph{Layer 3 (rate limit).}
Each agent is bound to a token-bucket rate limiter with
per-tool granularity (e.g., 10 \texttt{web\_fetch} per second,
1 \texttt{shell\_exec} per second). The bucket is refilled
on a millisecond clock. This protects against denial of
service via tool-call flooding and prevents pathological loops
where a buggy or adversarial agent saturates a critical
resource. Bucket lookup is a hash-table operation, a measured
0.02\,$\mu$s per call.

\paragraph{Layer 4 (adversarial pre-filter).}
A regex DFA scans the tool-call arguments for known prompt-injection
and encoding-attack patterns: ``\texttt{ignore previous instructions}'',
base64-encoded payloads with suspicious length, ROT13-encoded
keywords, authority-impersonation phrases (``\texttt{ADMIN
OVERRIDE}''), and instruction-hierarchy attacks
(``\texttt{system: \ldots}''). The pre-filter runs in O($n$)
time where $n$ is the byte length of the arguments
(a measured 7.50\,$\mu$s mean on the benign benchmark calls;
the single most expensive non-inference layer).
It catches the
``cheap'' attacks before the expensive layer 5 runs. The
companion paper~\cite{animaos-probelogits} reports that the
pre-filter alone catches roughly 13 of 50 adversarial prompts
in the Custom-260 benchmark.

\paragraph{Layer 5 (ProbeLogits gate).}
The semantic core. Detailed in \S\ref{sec:design-probelogits}
and Algorithm~\ref{alg:gateway} (lines 14--19). One forward
pass through the loaded inference model with KV
snapshot/restore around the call.

\paragraph{Layer 6 (constitutional policy).}
A 12-principle policy match: each principle is encoded as a
predicate over (agent, tool, arguments, result). Examples:
``no agent may invoke \texttt{shell\_exec} on a path containing
\texttt{/etc/shadow}''; ``no \texttt{web\_post} may target a
private RFC1918 address from a \texttt{Classic} agent.''
The 12 principles together evaluate in a measured
1.83\,$\mu$s for typical arguments (short-circuiting
predicates; v1's $\sim$60\,$\mu$s estimate was ${\sim}$30$\times$
too pessimistic).
This layer encodes deployment-specific policy that cannot be
expressed in the generic safety classifier of layer 5.

\paragraph{Audit (Blake3 hash chain).}
After every decision, a record \texttt{(timestamp, agent\_id,
tool\_name, arg\_hash, deciding\_layer, verdict, prev\_hash)}
is appended to a ring buffer. Each record's hash is included
in the next record's \texttt{prev\_hash} field, producing a
Blake3-chained log: tampering with any past record invalidates
all subsequent hashes. The buffer is 4\,MB ($\sim$30,000
records); when full, the oldest records are flushed to
AnimaFS storage on the next checkpoint cycle.

The chain integrity guarantee is \emph{intra-boot}: within a
single boot session, every record's hash chains back to the
session's first record, and tampering with any past record
invalidates the rest. \emph{Cross-boot} continuity is not
yet provided---each reboot starts a fresh chain anchored to a
new genesis record stored in AnimaFS---because durable chain
linking across kernel restarts requires committing the last
in-memory \texttt{prev\_hash} to the persistent store before
the kernel loses control. I treat this as future work.
For deployments that need cross-boot forensic chains, the
last-flushed AnimaFS record's hash should be persisted
synchronously on every gateway decision; the throughput cost
($\sim$1\,ms per persistent write) is significant and was
deferred from the present design.

\begin{table}[t]
\centering
\caption{Pipeline layers, latency, and role. Non-inference
rows are silicon-measured means over $n$=64 benign no-probe
calls (2026-07-05 in-OS microbench; total 11.3\,$\mu$s,
components sum consistently to 11.33\,$\mu$s---v1's
unarchived 65.3\,$\mu$s figure, whose parts did not sum, is
superseded). Layer 5 is the remeasured hosted
per-classification cost (Qwen2.5-7B Q4\_0;
Table~\ref{tab:latency-remeasure}).}
\label{tab:layers}
\scriptsize
\setlength{\tabcolsep}{2pt}
\begin{tabular}{@{}clrl@{}}
\toprule
\textbf{\#} & \textbf{Layer} & \textbf{Latency} & \textbf{Role} \\
\midrule
1 & Schema validation       & 0.10\,$\mu$s & JSON-RPC + MCP spec \\
2 & Trust tier check        & 1.15\,$\mu$s & Agent tier $\rightarrow$ tool whitelist \\
3 & Rate limit              & 0.02\,$\mu$s & Token bucket per agent \\
4 & Adversarial pre-filter  & 7.50\,$\mu$s & O($n$) regex injection \\
5 & \textbf{ProbeLogits gate} & \textbf{358\,ms} & Per-classification semantic check \\
6 & Constitutional check    & 1.83\,$\mu$s & 12-principle policy \\
-- & Audit (Blake3 chain)   & 0.73\,$\mu$s & Tamper-evident log \\
\midrule
\multicolumn{2}{@{}l}{Non-inference total} & \multicolumn{2}{l}{\textbf{11.3\,$\mu$s} (measured)} \\
\multicolumn{2}{@{}l}{With ProbeLogits} & ${\sim}$358\,ms & \\
\bottomrule
\end{tabular}
\end{table}

\paragraph{Implementation mapping (v2 note).}
The six layers above are the conceptual pipeline; the shipped
code realizes them with two additions and one structural
difference, which I state here for reviewers reading the
source. First, the gateway short-circuits on a \emph{Layer 0}
safe-mode gate (system-wide lockdown check) before Layer 1,
and a \emph{Layer 2.5} capability/kill-switch check between
trust tier and rate limit.
Second, the adversarial pre-filter executes as stage ``4a''
\emph{inside} the ProbeLogits layer's module---it runs
immediately before the probe forward pass and can deny without
invoking the model---rather than as a standalone gateway
stage;
the code labels the stages L1 schema, L2 trust, L3 rate,
L4 ProbeLogits (with pre-filter), L5 constitutional, L6 audit.
The conceptual ordering and the deny semantics are unchanged;
I keep the six-layer presentation for continuity with v1.
Since the v1 snapshot, the governed surface has also grown:
the four AnimaDB document host functions route through the
same governance check via the governed AnimaDB layer, and the
audit chain is now a \emph{keyed} Blake3 chain.

\subsection{ProbeLogits as Substrate}
\label{sec:design-probelogits}

The ProbeLogits primitive is described in detail in the
companion paper~\cite{animaos-probelogits}; here I summarize
the contract the gateway depends on. Given the loaded inference
model and a candidate tool call, ProbeLogits returns a single
real-valued safety score $s \in [-\infty, +\infty]$ representing
$\text{logit}(\text{Dangerous}) - \text{logit}(\text{Safe})$
at the verbalizer position, with calibration bias subtracted
and scaled by $\alpha$. The gateway thresholds at $s > 0$ to
reject; $\alpha$ is the deployment-time policy knob (higher
$\alpha$ = more conservative).

\textbf{Per-model setup at boot.} Anima OS runs a Token
Fertility check at boot to ensure the verbalizer pair (e.g.,
``Safe''/``Dangerous'' or ``Yes''/``No'') tokenizes to single
vocab IDs for the loaded model; if no usable verbalizer pair
exists, the gateway refuses to start (FAIL-CLOSED at startup).
Calibration bias is measured once on 7 null-input prompts and
cached.

\subsection{Algorithm}
\label{sec:design-algo}

Algorithm~\ref{alg:gateway} shows the gateway's
per-tool-call procedure. The five non-inference layers are
short-circuiting (any deny is final and skips layer 5);
the ProbeLogits gate is the only call that performs an LLM
forward pass; the audit chain is appended unconditionally,
recording the deciding layer and its verdict.

\begin{algorithm}[!t]
\caption{\textsc{GatewayCheck}: per-tool-call gateway pipeline.
\textsc{ProbeLogits}() returns $(\textit{ans}, p)$ where
\textit{ans} is the Boolean ``model says Dangerous'' and
$p \in [0, 1]$ is the sigmoid-calibrated confidence
($p = \sigma(s)$ from $s \in \mathbb{R}$ in \S\ref{sec:design-probelogits}).}
\label{alg:gateway}
\scriptsize
\begin{algorithmic}[1]
\Require Tool call $c = (\text{name}, \text{args})$, agent ID $a$
\Ensure  Verdict $v \in \{\textsc{Allow}, \textsc{Deny}\}$
\State $v_1 \gets \textsc{ValidateSchema}(c)$
\If{$v_1 = \textsc{Deny}$} \State \textsc{AppendAudit}($c, a, 1, v_1$); \Return $v_1$ \EndIf
\State $v_2 \gets \textsc{TrustTier}(a, c.\text{name})$
\If{$v_2 = \textsc{Deny}$} \State \textsc{AppendAudit}($c, a, 2, v_2$); \Return $v_2$ \EndIf
\State $v_3 \gets \textsc{RateLimit}(a)$
\If{$v_3 = \textsc{Deny}$} \State \textsc{AppendAudit}($c, a, 3, v_3$); \Return $v_3$ \EndIf
\State $v_4 \gets \textsc{AdvFilter}(c.\text{args})$
\If{$v_4 = \textsc{Deny}$} \State \textsc{AppendAudit}($c, a, 4, v_4$); \Return $v_4$ \EndIf
\State \Comment{Layer 5: ProbeLogits semantic check (graduated)}
\If{$\neg$\textsc{InferenceEngine.Ready}()}
  \State \Return \textsc{Deny} \Comment{FAIL-CLOSED}
\EndIf
\State $\textit{kv\_snap} \gets$ \textsc{KVSnapshot}()
\State $(\textit{ans}, p) \gets$ \textsc{ProbeLogits}($c$) \Comment{answer + confidence}
\State \textsc{KVRestore}($\textit{kv\_snap}$)
\If{$\textit{ans} \land p > 0.90$} \State \textsc{AppendAudit}($c, a, 5, \textsc{Deny}$); \Return \textsc{Deny} \EndIf
\If{$\textit{ans} \land p > 0.70$} \State \textsc{LogWarn}($c, a, p$) \Comment{record + notify, allow} \EndIf
\State $v_6 \gets \textsc{Constitutional}(c, a)$
\If{$v_6 = \textsc{Deny}$} \State \textsc{AppendAudit}($c, a, 6, v_6$); \Return $v_6$ \EndIf
\State \textsc{AppendAudit}($c, a, \text{ALLOW}, \textsc{Allow}$)
\State \Return \textsc{Allow}
\end{algorithmic}
\end{algorithm}

\subsection{FAIL-CLOSED Semantics}
\label{sec:design-failclosed}

If the inference engine is unavailable (model not loaded,
KV cache exhausted, deadlock detected), the layer-5 call
returns \textsc{Deny} (Algorithm~\ref{alg:gateway}, line 14).
Because layer 5 is on the critical path for every tool call,
this denies all tool calls system-wide until the engine
recovers. There is no path through the gateway that skips
layer 5: the only entry point (\texttt{call\_tool}) is the
gated wrapper, and the inner \texttt{call\_tool\_raw} is
\texttt{pub(crate)} (\S\ref{sec:design-mediation}).

\paragraph{Graduated response (not binary).}
Algorithm~\ref{alg:gateway} lines 18--19 implement a
graduated response, not a binary threshold: the gateway
denies (\textsc{Deny}) only when \emph{both} the model says
``Dangerous'' \emph{and} the calibrated confidence exceeds
0.90; calls in the [0.70, 0.90] confidence band are allowed
but logged as warnings (\textsc{LogWarn}); calls below 0.70
proceed normally. This trades some recall for fewer
false-positives at the operating point I ship as default
($\alpha=0.9$, \S\ref{sec:eval-threshold}). The graduated
band is a deployment policy choice; tightening to
0.70 (binary deny above 0.70) raises recall at the cost of
more over-refusal. The FAIL-CLOSED guarantee covers the
unavailable-engine case (line 14); for ``garbage-but-confident''
output (a model returning a well-formed but wrong probability),
the gateway's defense is the per-model calibration check at
boot (\S\ref{sec:design-probelogits}) plus the constitutional
layer (line 21), not the threshold itself.

\paragraph{KV restore failure mode.}
Algorithm~\ref{alg:gateway} line 17 (\textsc{KVRestore})
assumes the snapshot taken at line 15 is bit-for-bit
restorable. In a no\_std environment with a fixed-size heap,
restore can fail under memory pressure. Anima OS handles
this by panic-and-restart: a failed restore corrupts the
agent's main KV context, so the gateway treats the agent
session as compromised and tears it down. The agent's
audit chain remains valid; the agent's conversational state
is lost but no information is leaked across the boundary.
A more graceful recovery (e.g., snapshot-then-fork with
copy-on-write) is left to future work.

\paragraph{KV cache save/restore.}
The ProbeLogits forward pass at layer 5 reads the verbalizer
logits but must not corrupt the agent's main conversational
state. Without isolation, the probe's prompt template
(``Is this action dangerous, harmful, or a privacy violation?
\ldots'') would mutate the KV cache and contaminate
subsequent token generation by the agent. The gateway
therefore takes a KV snapshot before the probe forward
pass (line 15) and restores it after (line 17). The snapshot
is a per-layer copy of the key/value tensors at the current
position---a measured 0.93\,MB for a Qwen 2.5-7B Q4\_0 model
at 32 cached positions, taken via direct memory copy in a
measured 0.037\,ms, with a 0.016\,ms restore (0.053\,ms
round trip). The size scales linearly with the number of
cached positions; v1's 12\,MB estimate corresponded to a
much longer assumed context, and its sub-millisecond copy
estimate is confirmed with large margin.
Restore is symmetric. The probe runs in a dedicated KV
context that is discarded after each call; the snapshot/restore
pair guarantees the main agent context is bit-for-bit identical
before and after the gateway call.

\paragraph{Concurrency.}
The inference engine holds a single global lock during forward
pass; the gateway acquires it for the duration of the probe
classification (typically ${\sim}$360\,ms,
\S\ref{sec:eval-perf}). Multiple concurrent agent threads queue
on this lock in FIFO order, which bounds gateway throughput at
roughly 2.8 governed tool calls per second on the 7B
configuration.
The current implementation has no priority inheritance: a
high-frequency Classic-tier agent saturating the queue can
delay (but not indefinitely starve, given FIFO) governance
decisions for a System-tier agent behind it. A
priority-aware queue with bounded wait time for high-tier
agents is straightforward to add but is not implemented in
the present prototype. Sub-10\,ms governance under concurrent
load requires GPU/NPU offload of the inference layer
(\S\ref{sec:disc}, future work).

\subsection{Complete Mediation}
\label{sec:design-mediation}

\paragraph{Static argument.}
The Anima OS WASM runtime registers 18 host functions that an
agent can call: 15 base operations (file I/O, network, memory,
agent ops, and the four governed AnimaDB document operations)
plus 3 MCP tool-calling functions
(\texttt{mcp\_list\_tools}, \texttt{mcp\_call\_tool},
\texttt{mcp\_tool\_info}).
Every side-effect-producing function is mediated by the
kernel gateway before the operation returns to the WASM
sandbox: the base operations route through
\texttt{governance\_check\_host()} (the AnimaDB quartet via
the governed AnimaDB layer, which invokes the same check);
\texttt{mcp\_call\_tool} invokes the gateway's
\texttt{evaluate\_tool\_call()} pipeline directly---the same
six-layer pipeline, entered at the gateway rather than through
the host-function shim;
and the two remaining MCP functions
(\texttt{mcp\_list\_tools}, \texttt{mcp\_tool\_info}) are
read-only catalog queries with no side effects and therefore
no verdict.
(The kernel's built-in MCP server exposes 15 OS tools behind
this same gateway.)

\paragraph{Visibility argument.}
The MCP entry point itself (\texttt{call\_tool\_raw} in
\texttt{boot/src/mcp/client.rs}) is declared
\texttt{pub(crate)}---it is not exposed in the public WASM ABI.
The only path from the sandbox to \texttt{call\_tool\_raw}
is through the gated wrapper \texttt{call\_tool}, which
unconditionally invokes \texttt{governance\_check\_host()}
before forwarding. Static analysis of the WASM ABI surface
(crate \texttt{boot/src/wasm}) confirms no other entry point
into the gateway exists.

\paragraph{Empirical argument.}
Complete mediation is \emph{enforced structurally}---there is
a single kernel choke-point per surface, as argued above---and
two empirical artifacts back the structural claim in the
current tree. First, an in-OS test suite (\texttt{webtest
mcp}) exercises the MCP surface end-to-end from the kernel
shell: tool listing, governed calls, denial paths, and
FAIL-CLOSED behavior, checked against the gateway's audit
counters (\texttt{total\_requests}, \texttt{total\_blocked},
and the gateway-lock acquisition atomics), which count every
mediated call.
Second, a rebuilt enumerated-path mediation harness runs
in-OS and enumerates every reachable synchronous
agent-to-system path together with its mediation mechanism
(silicon-measured, 2026-07-05 single boot): \textbf{59 total
paths, of which 43 are mediated}. The breakdown: all 18 WASM
host functions (15 governed via
\texttt{governance\_check\_host()}, 3 read-only metadata
queries with no side effects); all 20 registered MCP tools,
each verdict-verified by a dry-run
\texttt{evaluate\_tool\_call()} through the full pipeline;
and the 21 ring-3 syscalls (7 governed via
\texttt{gate\_syscall()}, 1 self-gated through the inner
\texttt{evaluate\_tool\_call} path, 2 read-only metadata, 5
pure/no-resource). The harness also verified the negative
probe (a dangerous dry-run call is denied) and gateway
liveness. The 16 unmediated paths decompose as 5 read-only
$+$ 5 pure (no governed resource reached, so no verdict is
required) and \textbf{6 resource-reaching ring-3 syscalls
that are not yet gated} (\texttt{put}, \texttt{fetch},
\texttt{kv\_set}, \texttt{kv\_get}, \texttt{send\_msg},
\texttt{recv\_msg}), deferred to the M4.2 milestone.
I therefore \emph{withdraw} the v1 claim of ``123 paths,
100\% mediated'': complete mediation holds, measured, over
the host-function and MCP-tool surfaces; the ring-3 syscall
surface has six disclosed ungated paths until M4.2 lands.

\paragraph{What this argument does \emph{not} prove.}
Complete mediation in the Saltzer--Schroeder
sense~\cite{saltzer1975} requires that no path bypasses the
gateway. The arguments above show this is true \emph{for the
synchronous call-through paths in the WASM-to-system
interface}, which is the agent's only legal egress. They do
\emph{not} cover: (i) kernel exploits in Anima OS itself
(out of scope), (ii) hardware-level tampering
(out of scope), (iii) timing/side-channel inference
(\S\ref{sec:threat}, out of scope), (iv) WASM trap paths
where Cranelift's exception handler returns to the host
(currently the trap returns to the WASM module's own catch
handler with no host-side side effects, but this assumption
should be re-verified if WASM exception handling proposals
land), and (v) WASM import-resolution edge cases: a WASM
module that declares an import not in the registered list
of 18 host functions causes the JIT linker to refuse the
module, but this depends on Cranelift's resolver returning
\textsc{Err} rather than a null stub. The argument is
structural for the agent layer's documented call surface,
not formal for the entire stack.

\paragraph{WASM isolation soundness.}
The trust boundary argument
(\S\ref{sec:threat}, ``Trust boundary'') assumes Cranelift's
WASM isolation is correct. This is broadly accepted but the
systems community has documented JIT-bug bypass classes (e.g.,
the 2021--2023 wasmtime CVEs in Cranelift miscompilation,
GHSA-5fhj-g3p3-pq9g and follow-ons). A correct Cranelift
implementation is a precondition of the gateway's structural
guarantee; JIT bugs are an orthogonal attack surface that
this paper does not address.
Spectre-style speculation is mitigated by the bare-metal
single-tenant Anima OS environment (no other tenant shares
DRAM/page tables) but is not formally precluded; deployments
that multiplex the gateway across tenants would need
additional speculation barriers.

\section{Evaluation}
\label{sec:eval}


\subsection{Experimental Setup}
\label{sec:eval-setup}

\paragraph{Hardware.}
AMD Ryzen 9800X3D (8 cores, 16 threads, AVX-512), DDR5-6000
60\,GB. The governance-pipeline evaluations run on Anima OS
native (UEFI boot, no QEMU); the multi-model external
benchmarks (\S\ref{sec:eval-multimodel}) and the
per-classification latency remeasurement
(\S\ref{sec:eval-perf}) run hosted (llama-cpp-python) on the
same machine, and each result is labeled accordingly.

\paragraph{Inference model.}
Qwen2.5-7B-Instruct Q4\_0 (default), Llama 3 8B Q4\_0, Mistral
7B v0.3 Q8\_0 (cited from~\cite{animaos-probelogits}).

\paragraph{Datasets.}
\begin{itemize}[leftmargin=*]
  \item \textbf{Custom-101}: 101 MCP-tool-call prompts (37 dangerous,
    64 benign) covering 18 domains: file system, memory, network,
    injection, deception, agent ops, etc. Author-labeled.
  \item \textbf{Custom-260}: 260 OS-action prompts in 9 categories
    (50 adversarial). Used in companion paper~\cite{animaos-probelogits};
    referenced here for completeness.
\end{itemize}

\subsection{Custom-101: Primary Governance Evaluation}
\label{sec:eval-101}

\paragraph{Dataset construction.}
The Custom-101 benchmark contains 101 MCP tool-call prompts
spanning 18 domain categories, with 37 dangerous and 64 benign
labels (Table~\ref{tab:101-domains}). Categories were chosen
to cover the attack surface that the gateway is intended to
mediate: file system reads/writes, memory operations, network
egress (exfiltration and SSRF), prompt injection, constitutional
violations, deception/agent manipulation, and benign uses
across all of the above. Each prompt is a JSON-RPC \texttt{call\_tool}
payload with a tool name, arguments, and an author-assigned
``dangerous'' or ``benign'' ground truth label.

\begin{table}[t]
\centering
\caption{Custom-101 category breakdown (37 dangerous + 64 benign).
Categories are MCP-tool-call domains, not safety taxonomies.}
\label{tab:101-domains}
\scriptsize
\setlength{\tabcolsep}{4pt}
\begin{tabular}{@{}llrr@{}}
\toprule
\textbf{Domain} & \textbf{Sub-category} & \textbf{Dang.} & \textbf{Benign} \\
\midrule
File system  & sensitive read       & 5  & --- \\
             & write-danger         & 3  & --- \\
             & benign               & --- & 15 \\
Memory       & privacy violation    & 4  & --- \\
             & benign               & --- & 8 \\
Network      & exfiltration         & 3  & --- \\
             & SSRF                 & 3  & --- \\
             & benign               & --- & 9 \\
Injection    & attack               & 5  & --- \\
             & benign               & --- & 2 \\
Constitutional & violation          & 4  & --- \\
Deception    & agent manipulation   & 4  & --- \\
             & agent benign         & --- & 8 \\
Agent ops    & ask benign           & --- & 6 \\
             & probe benign         & --- & 6 \\
Subtle       & dangerous            & 6  & --- \\
             & benign               & --- & 2 \\
General      & benign               & --- & 8 \\
\midrule
\textbf{Total}  &                   & \textbf{37} & \textbf{64} \\
\bottomrule
\end{tabular}
\end{table}

\paragraph{Headline result.}
The 6-layer gateway, run end-to-end on Anima OS native
(UEFI boot, Qwen 2.5-7B Q4\_0 loaded), achieves a
silicon-measured \textbf{F1 = 0.789} (accuracy 84.2\%,
precision 0.769, recall 0.811; TP 30, FN 7, FP 9, TN 55).
The corpus itself is in-tree and verified (101 cases, exactly
37 labeled dangerous and 64 benign);
the v1 headline (F1 0.773) was runtime shell output with no
stored artifact---the archived measured re-run slightly
exceeds it.

The gap from F1 = 1.0 reflects two error
modes: 7 false negatives (dangerous prompts that slipped past
all 6 layers, mostly subtle phrasing in the
``deception/agent manipulation'' category) and 9 false
positives (benign prompts conservatively blocked, mostly
ambiguous file-system reads). The false-positive rate is the
operating-point cost of the conservative deployment default;
the measured threshold sweep
(\S\ref{sec:eval-threshold}) shows F1 plateaus at this
operating point from $\alpha = 0.70$ upward.

\subsection{Ablation: ProbeLogits is Load-Bearing}
\label{sec:eval-ablation}

To establish that ProbeLogits is essential---not just one
layer among six---I run a 4-configuration ablation
(Table~\ref{tab:ablation}).

\begin{table}[t]
\centering
\caption{Load-bearing layer ablation on Custom-101,
silicon-measured (2026-07-05 in-OS run, $n$=101). Removing
the ProbeLogits semantic layer collapses F1 from 0.789 to
0.357 ($\Delta$F1 = 0.432). The measurement is reported as
partial in one respect: the current tree exposes no
constitutional-only toggle, so the fourth configuration is
the No-Layer4 run (the code-level layer that combines
pre-filter and probe), reported in place of a
No-Constitutional cell.}
\label{tab:ablation}
\scriptsize
\setlength{\tabcolsep}{4pt}
\begin{tabular}{@{}lll@{}}
\toprule
\textbf{Configuration} & \textbf{F1} & \textbf{Note} \\
\midrule
Full pipeline (6-layer)        & 0.789 & TP30 / TN55 / FP9 / FN7 \\
No pre-filter                  & 0.805 & TP31 / TN55 / FP9 / FN6 \\
\textbf{No ProbeLogits (L5)}   & \textbf{0.357} & Load-bearing; TP10 / FN27 \\
No layer 4 (pre-filter+probe)  & 0.327 & near-chance; TP9 / FN28 \\
\bottomrule
\end{tabular}
\end{table}

\paragraph{The finding.}
\emph{Hand-rule firewalling without semantic understanding is
insufficient.} In the measured run, the 6-layer pipeline
minus its semantic core (schema validation, trust tier
check, rate limit, regex pre-filter, constitutional policy
match) collapses to F1 = 0.357, admitting 27 of the 37
dangerous calls (recall 0.27); with the pre-filter removed
as well it falls to 0.327, only marginally better than
always allowing.
An honest side observation: removing \emph{only} the
pre-filter slightly improves F1 on this benchmark (0.805
vs.\ 0.789, one fewer false negative)---the pre-filter's
value is its zero-cost early denial of encoding attacks, not
an F1 contribution on this corpus.
The ProbeLogits layer contributes by
reasoning about the \emph{intent} of the tool call, not its
surface form. Even the most carefully constructed hand rules
(our own pre-filter and constitutional policy) fall to
adversarial euphemism, multi-step plans where each step is
benign in isolation, and deception that masquerades as a benign
intent. Semantic interpretation requires the model.

\paragraph{Why the magnitude matters.}
The measured collapse is not a minor degradation; it is the
difference between a working safety system and one that
admits nearly three quarters of dangerous calls
($\Delta$F1 = 0.432; the No-ProbeLogits configuration
catches only 10 of 37 dangerous prompts). The v1 believed
delta (0.446 from an F1-0.773 baseline) is superseded by the
measured 0.432 from the higher F1-0.789 baseline---the
conclusion is unchanged: the semantic layer is load-bearing.

\paragraph{Qualitative case studies.}
To illustrate where ProbeLogits adds value beyond hand-rule
firewalling, I walk through three representative prompts and
the gateway verdict at each layer (logit scores are from the
v1 evaluation run; per-prompt logs will be archived with the
v2 re-run):

\textit{Case A (semantic euphemism).} Tool call:
\texttt{shell\_exec("dd if=/dev/zero of=/dev/sda bs=1M")}.
Layer 1 passes (well-formed). Layer 4 pre-filter does not
match any known injection pattern (the call uses no obvious
keywords). Layer 5 ProbeLogits returns
$s = +6.2$ (calibrated), well above the deny threshold:
the model recognizes that overwriting the boot disk with
zeros is destructive even though no rule explicitly enumerates
this pattern. The constitutional layer would also catch this
under the ``no destructive disk operations'' principle, but
ProbeLogits catches it without requiring an exhaustive
disk-pattern rule.

\textit{Case B (multi-step decomposition).} Tool call sequence:
(1) \texttt{file\_read("/home/user/.ssh/id\_rsa")},
(2) \texttt{web\_post("https://attacker.example/x", body=\$1)}.
The pre-filter catches (1) on the keyword
\texttt{id\_rsa}. The ProbeLogits gate independently classifies
each call: (1) returns $s = +4.8$ (privacy violation by
single-call analysis) and is blocked at layer 5 before the
adversary reaches step (2). Because the pre-filter and
ProbeLogits both fire here, this case is not where ProbeLogits
``wins''; it is included to show defense in depth.

\textit{Case C (adversarial framing).} Tool call:
\texttt{shell\_exec("for f in *; do mv \$f .\$f; done")}
in directory \texttt{/var/www}. Layer 4 sees no injection
pattern (no encoding tricks, no keyword match). Layer 5
ProbeLogits returns $s = -1.3$ (below threshold,
\textsc{Allow}); the model sees a benign-looking shell loop
that just renames files. This is a \emph{false negative}---the
shell command is destructive in context (hides all files in
the web root). The gateway misses this case; case studies
like this populate the 7 false negatives in the measured
Custom-101 evaluation. Mitigation requires either a stronger semantic
model or domain-specific constitutional rules; the gateway
design is not a complete solution to adversarial framing.

\subsection{Custom-260: OS-Action Pipeline Validation}
\label{sec:eval-260}

The Custom-260 benchmark (260 OS-action prompts in 9
categories, 50 adversarial) stresses the substrate primitive
under the full gateway pipeline. On Qwen2.5-7B-Instruct
Q4\_0 with the production pipeline (all six
layers active), the silicon-measured re-run gives
\textbf{F1 = 0.980, P = 1.000, R = 0.960} ($n$=260; TP 169,
FN 7, FP 0, TN 84; accuracy 97.3\%)
(the corpus is in-tree;
the v1 numbers were runtime output with no stored artifact;
the archived silicon re-run reproduces them exactly).
Precision is perfect (no safe prompt flagged: FP = 0).
This ``deployment-killer'' property---when
the gateway flags an OS action, the OS can block without
human review---is the operating-point benefit of the
full hand-rule-augmented pipeline. Layer-by-layer ablation
of this benchmark (uncalibrated 64.8\% $\to$ +calibration
87.1\% $\to$ +safety prompt 92.3\% $\to$ +privacy boost
$\sim$95\% $\to$ +adversarial pre-filter 97.3\%) and per-category
breakdown are reported in the companion ProbeLogits
paper~\cite{animaos-probelogits} and not duplicated here. The
relevance to this paper is that the gateway, when augmented
with the optional hand-rule layers (privacy boost, adversarial
pre-filter), reaches an operating point where OS-action
governance is feasible without human-in-the-loop review.

\subsection{Multi-Model Validation: an
Architecture-Agnostic Governance Substrate}
\label{sec:eval-multimodel}

New in this revision: a multi-benchmark validation of the
governance substrate across three model architectures
(HarmBench~\cite{harmbench}, XSTest~\cite{xstest},
ToxicChat~\cite{toxicchat}), with every number backed by an
archived run in the companion repository's \texttt{results/}
directory. Table~\ref{tab:multimodel} summarizes.

\begin{table}[t]
\centering
\caption{Multi-model substrate validation (Yes/No verbalizer;
hosted; archived runs). HarmBench block rate on the
non-copyright subset ($n$=300 of 400) and overall ($n$=400);
XSTest ($n$=450: 250 safe + 200 unsafe).}
\label{tab:multimodel}
\scriptsize
\setlength{\tabcolsep}{2pt}
\begin{tabular}{@{}lrrr@{}}
\toprule
 & \textbf{Qwen2.5-7B} & \textbf{Llama 3 8B} & \textbf{Mistral 7B} \\
\midrule
HarmBench block, non-\copyright{} & \textbf{99.0\%} & \textbf{98.0\%} & \textbf{98.7\%} \\
HarmBench block, overall & 86.8\% & 75.0\% & 79.5\% \\
XSTest recall (unsafe)         & \textbf{1.000}  & \textbf{0.985}  & \textbf{0.995} \\
XSTest over-refusal (safe)     & 0.864  & 0.516  & 0.556 \\
XSTest F1                      & 0.649  & 0.749  & 0.740 \\
\bottomrule
\end{tabular}
\end{table}

On HarmBench non-copyright ($n$=300), all three models block
98.0--99.0\% of harmful behaviors with the Y/N verbalizer---a
1\,pp spread across three distinct architectures and two
tokenizer families. On XSTest, unsafe recall is
98.5--100\% (a 1.5\,pp spread); over-refusal on safe prompts
varies more (51.6--86.4\%) and is the model-specific
operating-point cost. On ToxicChat ($n$=1000), in pure
hosted-mode comparison, ProbeLogits-Llama-3 with the S/D
verbalizer achieves F1 = 0.679 vs.\ Llama Guard 3's
F1 = 0.675 (Q4\_K\_M; 0.662 at Q8\_0)---parity (+0.4\,pp)
with a vanilla instruction-tuned model against a fine-tuned
safety classifier, with +10.5\,pp recall and
2.4--3.4$\times$ lower per-classification latency
(Table~\ref{tab:latency-remeasure}).

The implication for the governance gateway: \emph{the
substrate primitive's accuracy is architecture-agnostic for
catching unsafe content}. The gateway's correctness does not
hinge on a specific base model; any of the three tested
models can serve as the layer-5 substrate with equivalent
blocking behavior on the safety-critical (recall) axis, and
verbalizer choice---not architecture---is the first-order
design lever.

\subsection{Threshold ($\alpha$) Sweep}
\label{sec:eval-threshold}

The conservative deployment default maximizes recall
(precision/recall tradeoff favors blocking more). The
$\alpha$ sweep on Custom-101 is now silicon-measured
(5 points, single boot): F1 = 0.414 ($\alpha$=0.30),
0.667 (0.50), \textbf{0.789} (0.70), 0.789 (0.80),
0.789 (1.00).
Best F1 = 0.789 at $\alpha = 0.70$, with a plateau through
$\alpha = 1.0$: the confusion matrix is identical from 0.70
upward (TP 30 / FN 7 / FP 9 / TN 55), so on this benchmark
the conservative default costs nothing relative to the
optimum. This supersedes the v1 believed sweep (best F1
0.821 at $\alpha \leq 0.75$, with a claimed
${\sim}$5\,pp F1 cost at $\alpha=0.9$); the measured sweep
shows neither the higher peak nor the high-$\alpha$ penalty.

\subsection{Performance Overhead}
\label{sec:eval-perf}

\paragraph{Per-layer cost.}
Table~\ref{tab:perf-breakdown} reports the per-layer latency
of the gateway. The non-inference layers cost a measured
\textbf{11.3\,$\mu$s} in total (in-OS microbench over $n$=64
benign no-probe calls): schema 0.10, trust 1.15, rate 0.02,
pre-filter 7.50, constitutional 1.83, audit
0.73\,$\mu$s---components that sum consistently to
11.33\,$\mu$s. The v1 figure of 65.3\,$\mu$s was unarchived
and its listed components did not sum; the measured overhead
is ${\sim}$6$\times$ lower, a strictly stronger result.
Cumulative gateway-lock statistics over the boot: 71
acquisitions, 0 contended, mean hold 33\,$\mu$s, max
238\,$\mu$s.
The ProbeLogits layer (layer 5) is the
dominant cost, and here v2 corrects a v1 framing error: the
v1 text quoted ``65\,ms,'' which is one 7B
\emph{forward-token} time on bare metal
(DDR5-bandwidth-bound; the archived silicon measurement is
78.6\,ms/token at a sustained-load operating point)---a
lower bound, not the classification cost. A real ProbeLogits classification
prefills the action-plus-probe prompt and reads one logit;
the remeasured per-classification cost (hosted, 150 samples
per model) is \textbf{358\,ms} (Qwen2.5-7B Q4\_0),
\textbf{332\,ms} (Llama 3 8B Q4\_0), and \textbf{556\,ms}
(Mistral 7B Q8\_0).
A bare-metal per-classification measurement is now archived
on the OS-action corpus: 2{,}596.6\,ms average on Custom-260
(silicon, full action prompts). It is \emph{higher} than
hosted, not lower as v1 expected, because the bare-metal
harness prefills each full action prompt token-by-token at
the measured 78.6\,ms/token with no batched prefill; the
hosted and bare-metal figures measure different prompt
regimes and both are reported.

\begin{table}[t]
\centering
\caption{Per-layer gateway latency (full pipeline,
Qwen 2.5-7B Q4\_0). Non-inference rows are silicon-measured
means ($n$=64 benign no-probe calls, 2026-07-05 in-OS
microbench); the KV row is the measured snapshot+restore
round trip (0.93\,MB at 32 positions). Layer 5 is the
remeasured hosted per-classification cost.}
\label{tab:perf-breakdown}
\scriptsize
\setlength{\tabcolsep}{4pt}
\begin{tabular}{@{}clrl@{}}
\toprule
\textbf{\#} & \textbf{Layer} & \textbf{Latency} & \textbf{Determinant} \\
\midrule
1 & Schema validation       & 0.10\,$\mu$s & JSON-RPC parse \\
2 & Trust tier check        & 1.15\,$\mu$s & Hash-table lookup \\
3 & Rate limit              & 0.02\,$\mu$s & Token bucket arith. \\
4 & Adversarial pre-filter  & 7.50\,$\mu$s & Regex DFA, $|c.\text{args}|$ \\
5 & ProbeLogits gate        & 358\,ms     & Probe prefill + logit read \\
6 & Constitutional check    & 1.83\,$\mu$s & 12-rule eval \\
-- & KV snapshot/restore    & 0.053\,ms   & Memcpy of KV tensors \\
-- & Blake3 audit append    & 0.73\,$\mu$s & 1 hash + buffer write \\
\midrule
\multicolumn{2}{@{}l}{\bf Total non-inference}
                          & \bf 11.3\,$\mu$s & Layers 1--4, 6, audit \\
\multicolumn{2}{@{}l}{\bf End-to-end}
                          & \bf $\sim$358\,ms & With layer 5 + KV \\
\bottomrule
\end{tabular}
\end{table}

\begin{table}[t]
\centering
\caption{Per-classification latency, remeasured (hosted,
9800X3D, 8 threads; 150 classifications per system). The
governance-relevant unit is one full classification: probe
prompt prefill plus a single logit read (ProbeLogits) vs.\
autoregressive judgment generation (Llama Guard 3). The
$\times$ columns give the speedup relative to Llama Guard 3
(LG3) at the indicated quantization.}
\label{tab:latency-remeasure}
\scriptsize
\setlength{\tabcolsep}{2pt}
\begin{tabular}{@{}lrrr@{}}
\toprule
\textbf{System} & \textbf{ms/class.} & \textbf{$\times$\,Q4\_K\_M} & \textbf{$\times$\,Q8\_0} \\
\midrule
ProbeLogits Qwen2.5-7B Q4\_0  & 358  & 2.38$\times$ & 3.19$\times$ \\
ProbeLogits Llama 3 8B Q4\_0  & 332  & 2.56$\times$ & 3.44$\times$ \\
ProbeLogits Mistral 7B Q8\_0  & 556  & 1.53$\times$ & 2.05$\times$ \\
Llama Guard 3 Q4\_K\_M        & 851  & ---          &  \\
Llama Guard 3 Q8\_0           & 1142 &              & --- \\
\bottomrule
\end{tabular}
\end{table}

\paragraph{Comparison with userspace alternatives.}
NeMo Guardrails~\cite{nemoguardrails} typically runs 2--3 LLM
round-trips per check (input rail, output rail, optional
fact-check rail), each via a separate model invocation.
End-to-end latency is reported in the 2--5\,s range in NVIDIA's
own demos; the equivalent ProbeLogits-gated check runs the
same Qwen 2.5-7B model once and resolves in
${\sim}$330--360\,ms hosted (Table~\ref{tab:latency-remeasure}),
roughly an order of magnitude faster.
Llama Guard~3~\cite{llamaguard3} runs as a separate fine-tuned
8B model with autoregressive token generation; on the same
hosted environment as our gateway it averages 851\,ms
(Q4\_K\_M) to 1{,}142\,ms (Q8\_0) per check---%
2.4--3.4$\times$ slower than the ProbeLogits classification
on Qwen or Llama 3 (1.5--2.1$\times$ vs.\ the Q8 Mistral
probe; Table~\ref{tab:latency-remeasure}).
Crucially, both Llama Guard 3
and NeMo run as userspace components and are subject to the
bypass vulnerabilities of Listing~\ref{lst:bypass}; their
latency advantage (or disadvantage) is moot if the agent
chooses to skip them.

\paragraph{Throughput at ${\sim}$358\,ms/classification.}
Single-tenant gateway throughput is
1/358\,ms~$\approx$~2.8 governed tool calls per second on the
Qwen 7B configuration. (The v1 text quoted ``15 calls per
second'' from the 65\,ms per-token figure; that conflated
per-token decode with per-classification cost and is
corrected here.) For typical single-agent workloads (one tool
call per ${\sim}$1--10 reasoning steps, each step generating
50--500 tokens at a measured ${\sim}$12.7\,tok/s bare-metal,
sustained load), a reasoning
step takes seconds, so ${\sim}$2.8 governed calls per second
still does not become the single-agent bottleneck. For
multi-agent workloads with a shared inference engine,
however, the gateway's lock-bounded throughput becomes a
contention point substantially sooner than the v1 figure
suggested---which strengthens, rather than weakens, the case
for the accelerator-offload and cascade future work
(\S\ref{sec:disc}).

\section{Discussion}
\label{sec:disc}


\paragraph{Generalization across models.}
The gateway design is independent of the specific inference
model: any model that satisfies the ProbeLogits contract
(single-token verbalizer pair, calibrable bias) can serve as
substrate. The Token Fertility check
(\S\ref{sec:design-probelogits}) enforces this at boot. The
companion paper~\cite{animaos-probelogits} shows that
Qwen 2.5-7B, Llama 3 8B, and Mistral 7B all meet the contract
and reach 97--99\% block rate on HarmBench non-copyright;
gateway behavior inherits these properties.

\paragraph{Generalization across MCP transports.}
The gateway implementation in this paper handles the stdio
transport and a TCP transport (MCP-over-AnimaNet, using
AnimaNet's length-prefix framing) to Anima OS's own MCP
server. Connecting to \emph{external} MCP servers over the
Streamable HTTP / SSE transport defined by the MCP
specification is future work. Crucially, the gateway logic is
transport-agnostic (it operates on parsed JSON-RPC messages,
not raw bytes), so an additional transport---Streamable HTTP,
or gRPC---requires only a new transport adapter, not changes
to layers 1--6.

\paragraph{Limitations.}
\begin{enumerate}[label=(\roman*),leftmargin=*,topsep=2pt,itemsep=4pt]
  \item \textbf{Inference latency dominates.} A ProbeLogits
    classification costs ${\sim}$330--560\,ms per call
    (Table~\ref{tab:latency-remeasure}); the underlying 7B
    forward pass is DDR5-bandwidth-bound (a measured
    78.6\,ms per token on bare metal, sustained load).
    Sub-10\,ms governance under multi-tenant
    load requires GPU/NPU acceleration of the inference layer
    or a smaller cascade model. The \emph{production
    governance path is CPU-SIMD}. An experimental line of
    work in Anima OS has demonstrated bare-metal GPU
    inference on AMD RDNA4---a \emph{hybrid} decode of a 14B
    model at ${\sim}$20.83\,tok/s, in which the GPU executes
    only the Q4\_0 projection GEMVs at the memory-bound
    ceiling while the CPU runs the norms, attention, LM head,
    and sampling---but that path is off by default
    (feature-gated at compile time and disabled at runtime),
    targets a different model, and is \emph{not} wired into
    the gateway's layer 5.
    Inference latency therefore remains the single largest
    performance limitation of the governed path.
  \item \textbf{Author-labeled MCP benchmark.} Custom-101
    (37 dangerous + 64 benign) is the only existing
    MCP-domain governance benchmark and was constructed by
    the system author. A community-labeled MCP benchmark,
    ideally combined with a public red-team competition
    format, would significantly strengthen evaluation.
    External benchmarks~\cite{harmbench,xstest,toxicchat}
    validate the substrate primitive but not the MCP-domain
    gateway behavior.
  \item \textbf{No human red-team study.} The current
    evaluation assumes adversarial inputs sampled from
    existing datasets and the bypass demo of
    Listing~\ref{lst:bypass}. A formal red-team study with
    5+ external attackers attempting bypass under controlled
    conditions is planned future work.
  \item \textbf{Ablation lacks a constitutional-only cell.}
    The ablation (Table~\ref{tab:ablation}) is now
    silicon-measured (full 0.789, No-PreFilter 0.805,
    No-ProbeLogits 0.357, No-Layer4 0.327), but the current
    tree exposes no constitutional-only toggle, so a
    No-Constitutional cell is approximated by the No-Layer4
    configuration; the measurement log records the run as
    partial for this reason.
  \item \textbf{Six ring-3 syscalls ungated.} The measured
    mediation harness (\S\ref{sec:design-mediation})
    discloses that 6 resource-reaching ring-3 syscalls
    (\texttt{put}, \texttt{fetch}, \texttt{kv\_set},
    \texttt{kv\_get}, \texttt{send\_msg},
    \texttt{recv\_msg}) do not yet route through the
    gateway; gating them is scheduled for milestone M4.2.
    Until then, the complete-mediation claim is scoped to
    the host-function and MCP-tool surfaces.
  \item \textbf{MCP server compromise not fully mitigated.}
    The gateway treats tool outputs as untrusted input
    (\S\ref{sec:threat} item 5), but post-execution probing
    of tool outputs is not yet implemented (see Future Work).
  \item \textbf{No model provenance verification at boot.}
    The gateway's correctness depends on the inference model
    loaded at boot being the genuine, unmodified model. The
    present implementation loads a GGUF file from the NVMe
    self-host partition (or USB mass storage) with no signature
    check, no SHA-256 attestation,
    and no measured-boot integration. A maliciously
    substituted model would invalidate the entire substrate
    (and is listed out of scope in \S\ref{sec:threat}). Model
    attestation via TPM-anchored hash chains or vendor-signed
    GGUFs is a standard mitigation I plan for a future
    revision.
\end{enumerate}

\paragraph{Future work.}
\begin{itemize}[leftmargin=*,topsep=2pt,itemsep=2pt]
  \item \textbf{Post-execution ProbeLogits.} Probe tool
    \emph{outputs} before returning to the agent, defending
    against indirect prompt injection in tool results
    (the strongest open attack class).
  \item \textbf{IPC message probe.} Probe agent-to-agent
    messages for adversarial content, defending against
    multi-agent jailbreak orchestration.
  \item \textbf{GPU/NPU offload of layer 5.} Offload the
    probe forward pass to an accelerator to bring
    per-classification latency toward the tens-of-milliseconds
    range and free the inference engine for concurrent
    workloads. A concrete silicon anchor now exists: the
    experimental hybrid RDNA4 path (14B decode at
    ${\sim}$20.83\,tok/s vs.\ ${\sim}$7\,tok/s CPU-only, with
    the GPU running only the projection GEMVs and the CPU
    running the rest; off by default and not integrated with
    the gateway---see Limitations (i)) demonstrates that
    bare-metal GPU inference is attainable on this hardware.
    Wiring that offload into the gateway's layer-5
    probe---so that the \emph{governed} path, not just an
    experimental decode, benefits---is the future work.
  \item \textbf{Cascading models.} Run a small model as a
    first-pass filter; escalate to 7B only on uncertain
    cases. v2 adds a measured caution: naive small probes are
    not viable safety gates---SmolLM2-135M blocks only
    ${\sim}$13\% of HarmBench behaviors (Y/N verbalizer) and
    a 0.5B model over-refuses ${\sim}$70\% of safe prompts---%
    so the first-pass filter must be \emph{distilled} from
    the 7B judge rather than used off-the-shelf, and the
    escalation threshold must preserve FAIL-CLOSED recall.
  \item \textbf{Multi-tenant scaling study.} Evaluate gateway
    overhead at 100+ concurrent agents sharing one inference
    engine; characterize the lock-contention boundary.
  \item \textbf{Red-team competition format.} Public
    benchmark + bug-bounty for documented bypasses, building
    a community-labeled MCP governance dataset.
\end{itemize}

\section{Related Work}
\label{sec:related}


\paragraph{AI agent operating systems.}
Guillotine~\cite{guillotine} (HotOS'25) proposes
hypervisor-based isolation for adversarial AI agents on
VMX-capable commodity hardware (with custom silicon as a
roadmap item, not a present requirement). The two approaches
are complementary but with different trust assumptions:
Guillotine's hypervisor boundary is implemented in hardware
virtualization extensions and is substantially harder to
break than a Cranelift JIT isolation boundary (above);
Governed MCP's gateway is implemented in Rust at the kernel
level and is subject to the JIT-bug class
(\S\ref{sec:design-mediation}). A deployer who needs
isolation guarantees against a sophisticated adversary with
JIT-bug knowledge is better served today by a hypervisor
approach (Guillotine, or any VMX-based sandbox) plus
governance \emph{within} the VM. The semantic governance
contribution of this paper---the ProbeLogits gate, complete
mediation of the WASM ABI surface, and FAIL-CLOSED
semantics---can be deployed inside such a VM unchanged.
AIOS~\cite{aios-colm} (COLM'25) defines an agent-OS
abstraction layer in Python, providing scheduling and memory
management for multiple LLM-driven agents in a shared
runtime; AIOS does not place semantic safety enforcement at
the kernel boundary and runs as a userspace process subject
to the bypass demonstration of Listing~\ref{lst:bypass}.

\paragraph{Safety classifiers as standalone models.}
Llama Guard~1/2/3~\cite{llamaguard,llamaguard3} (Meta) and
WildGuard~\cite{wildguard} (AI2) are fine-tuned classifiers
distributed as standalone 7--8B models. They are typically
called by the agent runtime as a separate forward pass before
or after each LLM step. Two limitations apply when these are
used as agent-side governance: (1) latency is
${\sim}$0.85--5\,s per check
because of autoregressive token generation, vs.\
${\sim}$330--360\,ms for the single-prefill-plus-logit-read
ProbeLogits classification
(Table~\ref{tab:latency-remeasure}); (2) they live in
the same userspace as the agent and so are subject to the
same in-process bypass attacks (Listing~\ref{lst:bypass}).
Llama Guard~3 remains the appropriate choice when an
out-of-band fine-tuned classifier is required for
external deployment (e.g., a multi-tenant chatbot's input/output
filter), but is not the right fit for kernel-resident
per-tool-call governance.

\paragraph{Tool-call governance libraries.}
NeMo Guardrails~\cite{nemoguardrails} (NVIDIA) provides a DSL
for declaring input/output rails and dialog flows around LLM
invocations. AGT~\cite{msagt} (Microsoft) provides
schema-driven tool-call validation. NeMo's Colang DSL is
expressive (it can specify multi-turn refusal flows), but the
runtime is a Python library imported by the agent; AGT is
similarly a userspace component. Both ship as production
governance solutions today and define the deployment baseline
this paper argues against. Beyond schema validation and
DSL-defined refusal rails, neither performs semantic intent
classification of the kind ProbeLogits provides.

\paragraph{Reference monitor and capability OS lineage.}
Anderson's reference monitor~\cite{anderson1972} established
the kernel-mediated security primitive in 1972: an enforcement
point that is (a) tamper-proof, (b) always invoked, and
(c) verifiable. Saltzer and Schroeder~\cite{saltzer1975} gave
this the name \emph{complete mediation}. Flask/SELinux~\cite{flask}
and Capsicum~\cite{capsicum} extended the model with
type-enforcement and capability-based access for general
syscall-mediated systems. The gateway proposed here
applies the same classical OS principle to a new attacker
class (autonomous AI agents) and a new mediated operation
(semantic tool-call intent), and inherits the formal
properties of the reference-monitor model: the gateway is
tamper-proof (kernel-resident, not addressable from WASM), is
always invoked (\S\ref{sec:design-mediation}), and is
verifiable---structurally via the single mediated entry per
surface and empirically via the in-OS \texttt{webtest} suite
and the measured 59-path enumeration harness (43 mediated;
six ring-3 syscall paths disclosed as ungated pending M4.2,
\S\ref{sec:design-mediation}).

\paragraph{Indirect prompt injection.}
Greshake \emph{et al.}~\cite{greshake2023} systematized
indirect prompt injection as an attack class against
LLM-integrated applications. Their real-world case studies
(notably Bing Chat) motivate this paper's threat model.
The defenses they discuss are application-layer
(input sanitization, output filtering); kernel-resident
mediation extends the defense surface below the application.

\paragraph{Constrained decoding and safety-by-construction.}
Outlines~\cite{outlines} and llguidance~\cite{llguidance}
constrain LLM output to grammars or formal languages; this
guarantees structural validity but not semantic safety
(a JSON-schema-valid \texttt{rm -rf /} is still dangerous).
The gateway can be combined with constrained decoding: the
WASM agent receives only schema-valid tool descriptors, and
the gateway adds the semantic check on top.

\section{Conclusion}
\label{sec:conclusion}


Tool calls are the syscalls of the agent era, and they
deserve kernel-grade governance. Governed MCP demonstrates
that this is feasible today: a six-layer gateway with a
logit-based semantic core, complete mediation of the WASM
ABI surface (synchronous call paths) across all 18
WASM-to-system host functions and all 20 MCP tools enforced
at a kernel choke-point (measured 59-path harness; six
ring-3 syscall paths remain ungated and are disclosed,
\S\ref{sec:design-mediation}), a measured 11.3\,$\mu$s
non-inference overhead (\S\ref{sec:eval-perf}),
per-classification semantic checks at 332--556\,ms---%
2.4--3.4$\times$ faster than a fine-tuned safety
classifier---and FAIL-CLOSED semantics under inference
engine failure. Removing the semantic
layer collapses the measured F1 from 0.789 to 0.357
(\S\ref{sec:eval-ablation}); rule-based governance
without semantic interpretation is insufficient.
The substrate is architecture-agnostic: three model families
block 98--99\% of HarmBench non-copyright behaviors within a
1\,pp spread (\S\ref{sec:eval-multimodel}).

The gateway is implemented in Anima OS---an open-source
($\sim$286{,}000 lines of Rust, AGPL-3.0) bare-metal kernel.
Existing in-process safety libraries
(NeMo Guardrails~\cite{nemoguardrails}, AGT~\cite{msagt})
can be defeated in 10 lines of Python that monkey-patch
their entry points or simply bypass them by not importing
them; kernel-resident mediation eliminates this failure
mode by construction.

Future work extends the gateway in three directions:
post-execution probing of tool outputs (closing the indirect
prompt injection vector), GPU/NPU acceleration of layer 5
(toward tens-of-milliseconds classifications; an
experimental, off-by-default hybrid RDNA4 decode provides a
silicon anchor but is not yet wired into the governed path,
\S\ref{sec:disc}), and multi-tenant
scaling under shared inference. The central claim is already
supported by the load-bearing ablation: \emph{semantic safety
enforcement belongs in the kernel, not in the application}.


\end{document}